\input{aipcheck}

\documentclass[  ]
  {aipproc}

\layoutstyle{6x9}

\begin{document}

\def\msun{\rm M_{\odot}}
\def\etal{{et al.\ }}
\def\simlt{\mathrel{\rlap{\lower 3pt\hbox{$\sim$}}\raise 2.0pt\hbox{$<$}}}
\def\simgt{\mathrel{\rlap{\lower 3pt\hbox{$\sim$}} \raise 2.0pt\hbox{$>$}}}
\def\lsim{\mathrel{\rlap{\lower 3pt\hbox{$\sim$}}\raise 2.0pt\hbox{$<$}}}
\def\gsim{\mathrel{\rlap{\lower 3pt\hbox{$\sim$}} \raise 2.0pt\hbox{$>$}}}
\def\di{\mbox{d}}
\def\mbulge{M_{\rm Bulge}}
\def\msunpc3{\msun~{\rm {pc^{-3}}}}
\newcommand{\be}{\begin{equation}}
\newcommand{\ee}{\end{equation}}
\def\kms{{\rm\,km\,s^{-1}}}

\title{On the inspiral of massive black holes\break in gas-rich galaxy mergers}

\classification{PACS Number 98.54}
\keywords      {Black hole physics -- hydrodynamics --  galaxies: starburst -- galaxies: evolution -- galaxies: nuclei}

\author{M. Colpi}{
  address={Dipartimento di Fisica G. Occhialini, Universit\`a degli
  Studi di Milano Bicocca, Piazza della Scienza 3, 20126 Milano, Italy}
}  
\author{S. Callegari}{
   address={Institute for Theoretical Physics, University of Z\"urich,
  Winterthurerstrasse 190, 
  CH-8057 Z\"urich, Switzerland}
}

\author{M. Dotti}{
  address={Dipartimento di Fisica e Matematica, Universit\`a
  dell'Insubria, Via Valleggio 11, 22100 Como, Italy}
}

\author{S. Kazantzidis}{
  address={Kavli Institute for Particle Astrophysics and Cosmology,
  Department of Physics, Stanford University, MS 29, Stanford,
  California 94309, USA}
}

\author{L. Mayer}{
   address={Institute for Theoretical Physics, University of Z\"urich,
  Winterthurerstrasse 190,
  CH-8057 Z\"urich, Switzerland}
  ,altaddress={Institute of Astronomy, ETH Z\"urich-H\"onggerberg,
  CH-8093 Z\"urich, Switzerland} 
}

\begin{abstract}
We present a study on the dynamics of massive black holes (BHs) in
galaxy mergers, obtained from a series of high-resolution N-Body/SPH
simulations.  We show that the presence of {\it a gaseous component}
is essential for the {\it rapid} formation of an eccentric (Keplerian)
BH binary.  The binary resides at the center of a massive ($\sim
10^9\msun$) turbulent nuclear disc resulting from the collision of the
two gaseous discs present in the parent galaxies.  Using physically
and/or numerically motivated recipes, we follow the accretion history
of the BHs during the merger.  We find that the mass of the BHs
increases as large central inflows of gas occur inside each galaxy,
and that the mass ratio $q_{\rm BH}$ varies with time indicating that
the memory of its initial value may be lost.  Given the uncertainties
in the accretion recipes and the encountered strong degeneracy between
numerical resolution and physical assumptions, we suggest here three
possible paths followed by the galaxies and the BHs during a merger in
order to fulfill the $M_{\rm BH}$ versus $\sigma$ relation : {\it
Adjustment, Symbiosis}, and {\it BH Dominance}. In an extremely high
resolution run, we resolved the turbulent gas pattern down to parsec
scales, and found that BH feedback is expected to be effective near
the end of the merger.  We then trace the BH binary orbit down to a
scale of $0.1$ pc modeling the nuclear disc, resulting from the galaxy
collision, as an equilibrium Mestel disc composed either of gas, gas
and stars, or just stars. Under the action of dynamical friction
against the rotating gaseous and/or stellar background the orbit
circularizes. When this occurs, each BH is endowed with its own
small-size ($\ll 0.01$ pc) accretion disc comprising a few percent of
the BH mass.  Double AGN activity is expected to occur on an estimated
timescale of $\lsim 10$ Myrs, comparable to the inspiral time.  The
double nuclear point--like sources that may appear have typical
separations of $\lsim 10$ pc, and are likely to be embedded in the
still ongoing starburst.

\end{abstract}

\maketitle

\section{I. Galaxy collisions with Massive Black Holes}

In this section we explore the dynamics of supermassive black holes
(BHs) in gas-rich galaxy mergers resulting from high-resolution
N-Body/SPH simulations carried out with Gasoline \cite{gasoline}.  The
initial galaxy models comprise a dark matter halo, a disc with 10\% of
its total mass in the form of gas, a bulge and a BH treated as
softened particle; the reference galaxy reproduces the Milky Way with
its central black hole ($M_{\rm BH}=3\times 10^{6}\msun$).  Encounters
with mass ratio $q=M_2/M_1$ equal to 1 and 1/4 are explored, the
smaller galaxy ($M_2$) being a rescaled replica of the Milky Way like
reference model; the BH mass ratio $q_{\rm BH}$ is the same as $q$,
according to the Magorrian relation \cite{magorrian98}.  Atomic
cooling down to a floor temperature of 20,000 K, and star formation
are accounted for in the simulations (see \cite{stelios} for details).
  
First, the merger and BH inspiral is followed from 
the large scale $\sim$ 100 kpc  
down to $\sim 200$ pc 
which is our force resolution limit.
In a second run, we apply the technique of particle 
splitting  to follow the gas thermodynamics and the
massive BH inspiral 
down to a force resolution of 4 pc (see \cite{lucioproc}). 

\medskip
$\bullet$ {\bf BH DYNAMICS down to 200 pc:} In encounters between
gas--rich galaxies with $q=1$ (``major'' mergers) the BHs, dragged
together with their more massive bulges, lose orbital angular momentum
under the action of dynamical friction. After $\sim 5$ Gyr, the BHs
form a close pair. They orbit inside a massive ($\sim 10^9\msun$)
self-gravitating nuclear disc formed at the end of the merger
\cite{hernquist95}.  In gas-free encounters with $q=1$, the BHs sink
at the center of the remnant forming again a close pair with
separations solely limited by the force resolution.

By contrast, in encounters with $q=1/4$ (``minor'' mergers)
the final separation of the BHs depends sensitively on how the central 
structure of the merging galaxies is affected by gaseous dissipation.
When absent (i.e., in dry mergers), mass loss induced by tidal
forces disrupt the 
secondary, lighter galaxy, leaving the less massive
BH wandering at a distance of several kpc 
away from the center of the remnant. In contrast, gas cooling
facilitates 
the pairing of the two BHs (down to 200 pc) 
by increasing the resilience 
of the secondary galaxy to tidal disruption.

\medskip
$\bullet$ {\bf BH DYNAMICS down to 4 pc:}
In the case of a  major ($q=1$) coplanar merger, we continued
the simulation using the technique of particle splitting
to explore the BH dynamics on a scale of few parsecs \cite{lucioproc}. 
We find 
that {\it the inspiral of the two BHs proceeds 
down to a distance at which a close Keplerian binary forms}.
Their sinking is controlled fully by the gaseous component 
under the typical conditions of a starburst.  
The BHs are found to move on a mildly eccentric orbit inside
a $\sim 10^{9}\msun$ rotationally supported, turbulent 
nuclear disc of size $\sim 40$ pc \cite{lucio}.

\medskip
$\bullet$ {\bf ACCRETION and BH GROWTH:}
The simulations covered a dynamical range of 5 orders
of magnitude (in total) from
the scale of 100 kpc down to 4 pc, and only in the 
refined run the force limit 
was becoming comparable to the gravitational
sphere of influence of the BHs: for this reason,
we could not probe the flow pattern around 
each individual BH along the course of their evolution
with enough accuracy to assess the magnitude of the accretion rate
$\dot M_{\rm BH}$.
We thus applied different recipes for ${\dot {M}}_{\rm BH}$ in order
to extract information about the ``possible'' mass growth from the N-body/SPH  
run of a $q=1/4$ merger with gas cooling and star formation.

We first considered a standard Bondi Eddington limited accretion
recipe for $\dot M_{\rm BH}$.  The density of the gas around each
massive BH is computed within a spherical radius twice our force
softening (i.e., 200 pc in the less resolved simulations, and 4 pc in
the high resolution run).  As reference speed for the gas in the Bondi
formula we considered either the thermal sound speed in the vicinity
of the BHs ($\sim 15 \,\kms$), or we hypothesized a turbulent velocity
for the gas $\sim 60\,\kms$: such turbulent motions are not resolved
in the low resolution simulations and are thus simply considered as
they are observed in circumnuclear discs of merger remnants
\cite{dowsol}. Instead, in the high resolution simulation we measured
turbulent velocities of such magnitude for the gas particles in the
nuclear disc.  They arise as a result of the final supersonic
collision between the two gaseous galaxy cores and persist because of
the inefficient dissipation implied by the effective equation of state
of the nuclear gas subject to the starburst.

\begin{figure}
  \includegraphics[width=0.8\textwidth,angle=270]{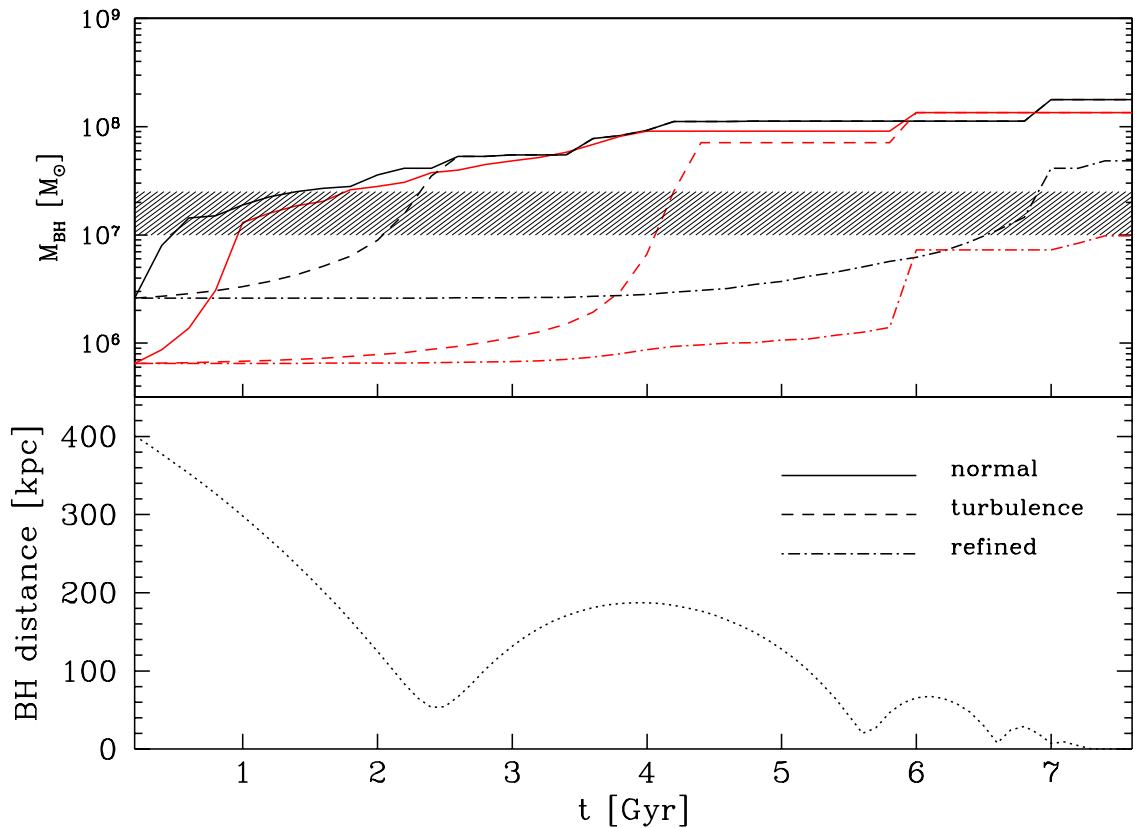} \caption{In
  the upper panel, different histories for the mass growth of the
  primary and secondary BHs are shown in black and red, respectively,
  for an encounter with $q_{\rm BH}=1/4$. Solid lines refer to
  standard Bondi-Hoyle accretion, dashed lines show results following
  from the assumption of turbulent motions, and dot-dashed lines trace
  the accretion history obtained extrapolating data from the refined
  simulation. In the lower panel, the relative distance of the two BHs
  during the merger is plotted.}
\end{figure}

Figure 1 shows in its lower panel the BH relative distance as a
function of time, and, in its upper panel, the 
corresponding mass growth of the BH. 
The black (red) solid line
describes the mass of the primary (secondary) BH as a function of
time, obtained from the low-resolution simulations using the thermal
sound speed in the Bondi formula.  Both BH masses grow initially not
because of gas inflows stemming from the merger process, but simply
because gas can be accreted inside a minimum volume fixed by the
numerical resolution (200 pc). This volume is too large to allow for a
realistic estimate of the gaseous mass effectively available (as seen
in similar numerical experiments, even for isolated galaxies; see
e.g. \cite{springel}).  Further accretion episodes occur at every
pericentric passage.  The long-dashed lines refer instead to accretion
obtained by considering, for the same simulation, a typical turbulent
sound speed. We see that mass growth occurs only after the tidal
disturbances, excited during the merger, drive gas inflows towards the
central regions of the interacting galaxies.  In both cases the BHs
become with time equally massive, and the memory of the initial ratio
$q_{\rm BH}$ is lost. This is due to the relative importance of tides,
that perturb more strongly the lighter galaxy.

At the end of the simulation, we extracted the value of the
line-of-sight stellar velocity dispersion of the merger remnant inside
1/8 of the effective radius (obtained by fitting the stellar mass
profile with a Sersic law \cite{sersic} and averaging the result over
different viewing angles).  We then inferred the mass that the final relic
BH would have in order to lie on the $M_{\rm BH}$ versus $\sigma$
relation (shown as the shaded area in Figure 1).  The two BH histories
discussed above imply masses that are in excess of the expected value
by an order of magnitude, even well before the merger is completed.
These histories imply that the BHs may self-regulate their
mass through AGN feedback 
on the
surroundings ``during the course of the merger'', sliding 
along the $M_{\rm BH}$ versus $\sigma$ relation.

The limitations related to the force resolution appear clear when
considering the third history (dot-dashed lines of Figure 1). In this
last case, we extracted information about the density profile and the
effective sound speed of the gas at a much closer distance (comparable
to the BH gravitational sphere of influence) by rescaling  
results from a refined high resolution simulation of 
equal mass galaxies.  These curves suggest that the BHs do not undergo
any major growth during the first two orbits, and only when the
galaxies merge the BH masses grow by more than an order of
magnitude. The BH mass ratio varies with time but does not seem to
increase toward unity.  The primary BH is approaching and exceeding
``from below'' the $M_{\rm BH}$ versus $\sigma$ expectation value, but
in this case the final discrepancy is consistent with the scatter of
the relation. The last episode of rapid mass growth may become
self-regulated once feedback is included.

\begin{figure}
  \includegraphics[width=.8\textwidth]{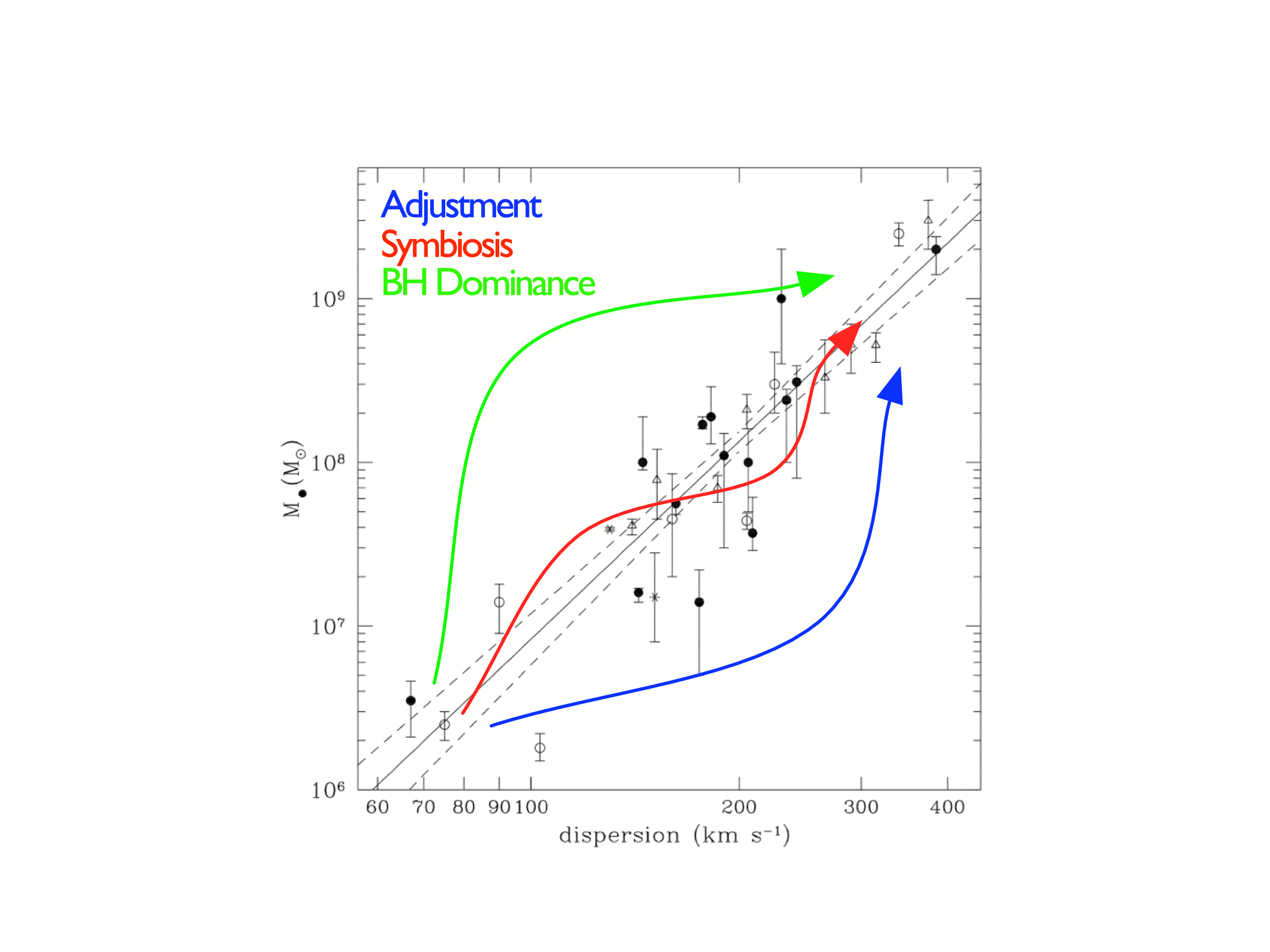}
  \caption{The three possible paths followed by galaxies and BHs
  during a merger in the BH mass versus stellar dispersion diagram from
Tremaine et al. 2002 \cite{tremaine}. 
The green arrow shows the trend for the {\it BH
  Dominance}  scenario, the red one refers to the {\it Symbiosis} scenario,
the 
  blue to {\it Adjustment}.} 
\end{figure}

Given all these uncertainties on the BH mass growth, we can depict
three possible (speculative) paths in the $M_{\rm BH}$ versus $\sigma$
relation. Figure 2 shows these possibilities: (1) {\it Adjustment}: Adjustment
occurs if
the mass growth time $\tau_{\rm growth}$ of the individual BHs is long
compared to the timescale of the merger $\tau_{\rm merger}$ (either
because of turbulence or excess of angular momentum content in the gas
near the massive BHs). The AGN phase would occur when the BHs are already
in place at the center of the remnant galaxy, and self-regulation 
would 
result from balance between (thermal and/or mechanical) energy
injection and gravitational energy of the surrounding gas. This can
happen either before or after the final coalescence of the BH binary.
In the first case, i.e., before
coalescence, a double nuclear pointlike source could be detected
in a star-bursting environment at separations closer than $\simeq 10$
pc.  (2) {\it Symbiosis}: In this case, 
both accretion and the buildup of
the final galaxy occur almost synchronously ($\tau_{\rm growth}\sim
\tau _{\rm merger}$) with step-like episodes of AGN activity and
starburst, leading to a ``sliding'' of $M_{\rm BH}$ and $\sigma$ along the
observed relation. In this case we may observe single or double AGN
activity. If double, this would occur at separations from several kpc
down to subparsec scales.  (3) {\it BH Dominance}: BH growth is
triggered by tidal torques during the early phase of the merger
($\tau_{\rm growth}< \tau _{\rm merger}$), resulting in a mass
exceeding the relation at intermediate stages; in this case the
assembly of the galaxy would be dictated by the BH (BHs) that may
regulate through its (their) luminosity the last starburst and the shape of
the final potential well determining the $\sigma$ of the remnant,
though it is unclear how this could produce the higher velocity
dispersion needed to satisfy the relation.  In this scenario, luminous 
double AGN activity would be observed well before the two galaxies lose their
identity, on very large scales.

Most theoretical efforts have been focused on the first scenario 
\cite{reessilk}, \cite{king} (see also \cite{granato}), 
and observations seem to find a low fraction
of widely separated, interacting galaxies hosting nuclear activity,
thus disfavoring -- albeit with some uncertainties -- BH dominance as
a general trend. On the other hand, numerical experiments on the scale
of mergers with force resolution in the commonly employed range (at
least an order of magnitude larger than the sphere of influence of a
BH) indicate the third case as possible, and avoid/tune excessive BH
mass growth by invoking a certain amount of local AGN feedback acting
on the surrounding gas \cite{springel}. However, assessing with enough
confidence which one of the three paths is actually followed would
require convergence of results when going to higher resolution. 
Instead, the striking differences in the BH mass accretions shown in
Figure 1 suggest that there is a strong degeneracy between numerical
resolution and physical assumptions. In the absence of any
feedback from the BH itself, higher numerical resolution produces the
same delay in the growth of BHs as unresolved turbulence. 
The high resolution simulations indicate that most of the BH growth
occurs after the merger, which favors the "adjustment" scenario.

\section{II. Black hole dynamics in rotationally supported nuclear discs}

The last, independent series of simulations (carried out with Gadget;
see \cite{gadget}) trace the dynamics of a BH pair (with $q_{\rm
BH}=1, 1/4, 1/10$) orbiting inside a self-gravitating, rotationally
supported disc (of $10^8\msun$) composed either of gas, gas and stars,
or just stars \cite{dotti1},\cite{dotti2}. 
A primary BH (of mass $4\times 10^6 \,\msun$) is placed
at the center of the disc, while a secondary is set on an eccentric
orbit, according to the outcome of the large--scale simulation (see
\cite{lucioproc} and Part I).  The disc has a radial Mestel profile,
and a Toomre parameter $>3$ everywhere, to ensure stability.
 
Figure 3 shows the BH relative separation and orbital eccentricity 
as a function of time for $q_{\rm BH}=1/4$ and a star-to-total mass ratio
$f_*$ in the disc of 0, 1/3, 2/3, and 1, respectively. We find that
dynamical friction against the gaseous and/or stellar background is
responsible for the inspiral of the BH along the full orbit, and
that the response of the background is insensitive to $f_*$, as the
orbital motion of the secondary BH is highly supersonic \cite{ostriker}. 

\begin{figure}
  \includegraphics[width=0.8\textwidth]{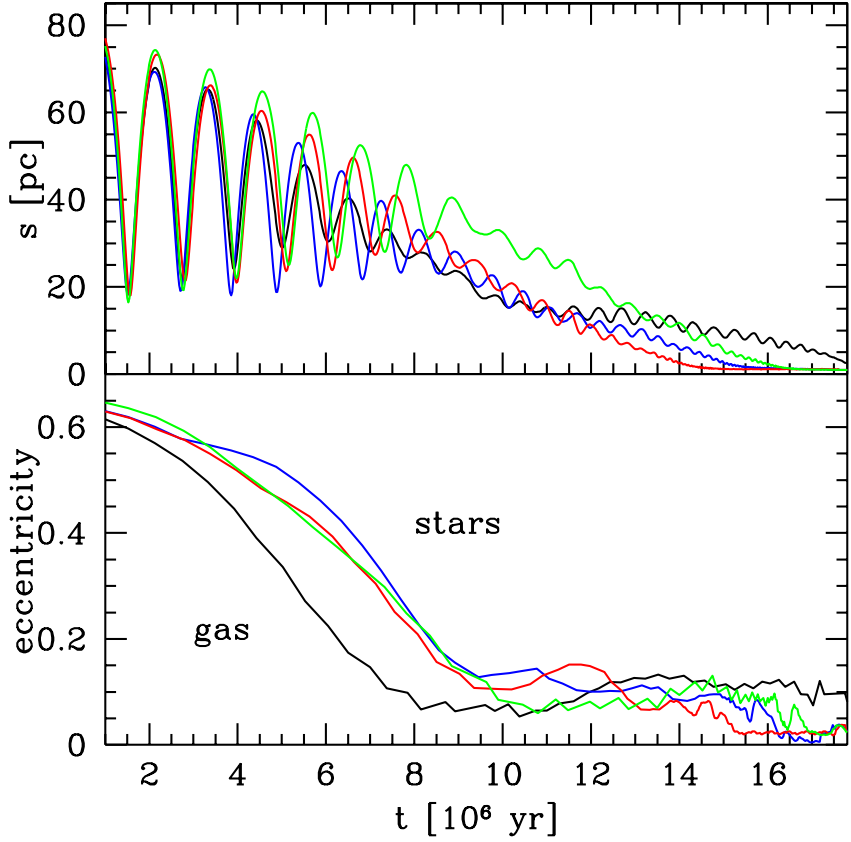}
  \caption{BH mass ratio $1/4$.  
  Upper panel: separations $s$ (pc) between the BHs as a
  function of time. Lower panel: eccentricity of the BH binary as a
  function of time. Black, blue, red and green lines refer to stellar
  to total disc mass ratio $f_*$ of 0, 1/3, 2/3 and 1. 
The two BHs form a binary for separations $s<5$ pc.}
\end{figure}

All co-rotating, initially eccentric orbits circularize, as angular
momentum loss by friction is less efficient than energy loss.  Figure
4 shows the density perturbation excited by the secondary BH in the
gas (left panels) and stellar (right panels) components at two
different times, corresponding to the first passage at pericenter
(upper panel), and apocenter (lower panel). The green curve shows the
counterclockwise corotating orbit of the BH. Near pericenter, the BH
has a speed larger than the local rotational velocity, so that
dynamical friction causes a reduction of the velocity of the BH: the
density wake lags behind the BH trail. On the contrary, near
apocenter, the BH speed is lower than the disc rotational velocity,
and, in this case, the force increases the orbital angular momentum
since the wake is dragged in front of the BH trail. The net effect is
a severe reduction of the eccentricity.

\begin{figure}
  \includegraphics[width=0.88\textwidth]{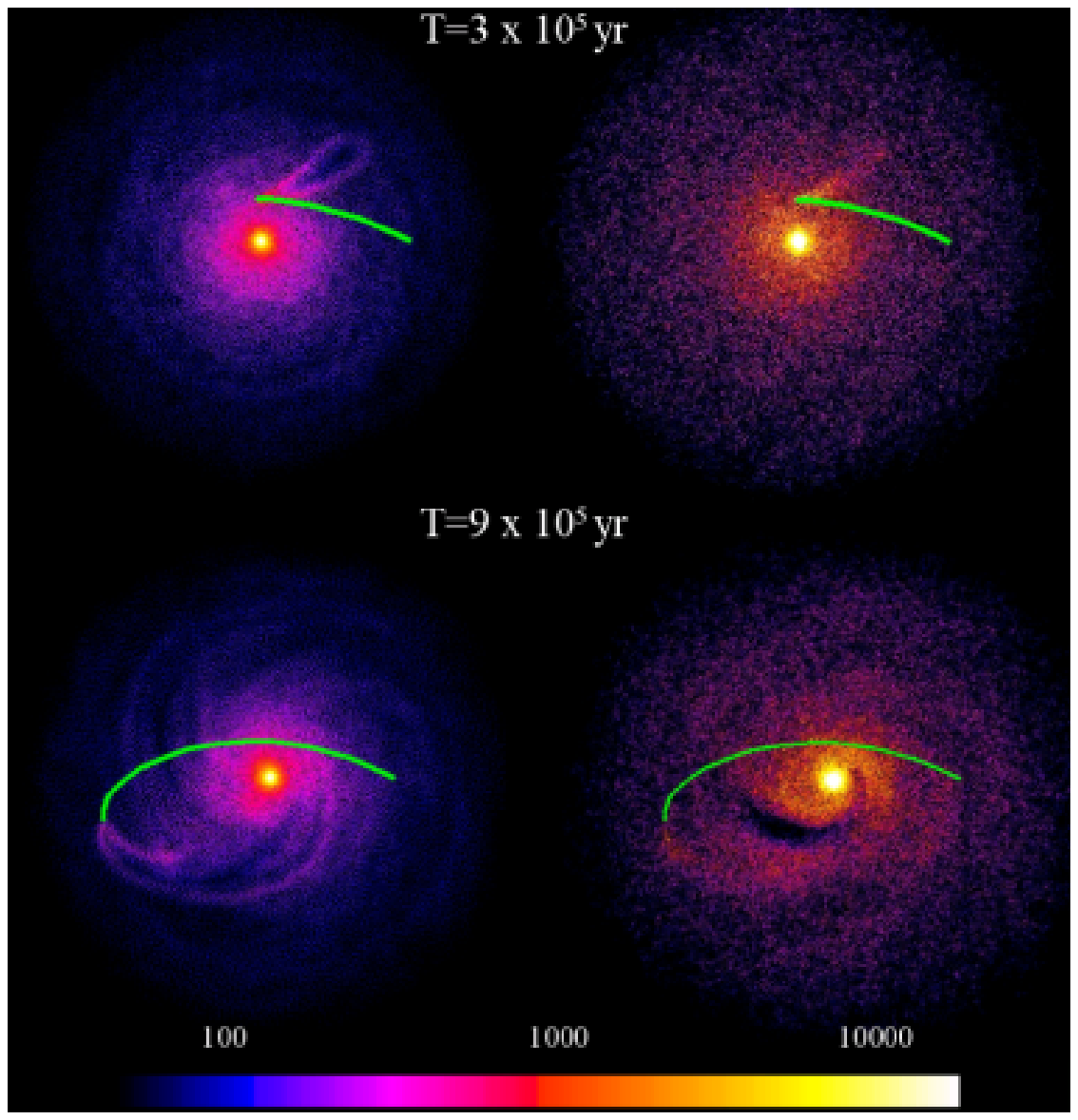} \caption{ Snapshot
  from a simulation with both gaseous ($2/3$ of the total disc mass)
  and stellar (1/3) disc components seen face--on.  The color coding
  indicates the z--averaged gas density (left panel) and star density
  (right panel) in logarithmic scale (units $\msun$/pc$^3$). The green
  lines trace the secondary BH counterclockwise prograde orbit.  The
  primary BH is present in the center of the disc. Each box size is
  200 pc.}
\end{figure}

In the purely gaseous run with $q_{\rm BH}=1,$ we spatially resolve,
on sub-parsec scales, the gravitational sphere of influence of each BH
(of the order of $\simeq 6$ pc) to detail the mass profile of bound
gas particles. We define as strongly (weakly) bound gas particles SBPs
(WBPs) those found inside the sphere of influence with total energy
per unit mass $E=0.5\Omega$ ($E\sim 0$), where $\Omega$ is the
gravitational potential of each BH.  We find that the mass collected
by each BH is $M_{\rm SBP}\approx 0.02 M_{\rm BH}\approx 8\times
10^4\msun$, and $M_{\rm WBP}\approx 0.85 M_{\rm BH}\approx 3.4\times
10^6\msun$, respectively.  These masses are of the same order for both
the primary and secondary BHs, and remain constant in time as long as
the BH separation is $\gsim 1$ pc. At shorter relative distances the
WBPs experience the tidal field of the binary so that they are no
longer bound to the individual BH.  At the end of the simulation
$M_{\rm WBP}$ is smaller, as shown in Figure 5.  At the same time the
mass $M_{\rm SBP}$ associated to the primary (secondary) BH increases
by a factor $\approx 4$ ($\approx 2.5$).  Bound particles have a net
angular momentum with respect to each BH, and form an ellipsoidal
configuration which is pressure supported. The typical half--mass
radii are similar for the two BHs: $\simeq 3$ pc and $\simeq 0.2$ pc
for the WBPs and SBPs, respectively.

\begin{figure}
  \includegraphics[width=0.8\textwidth]{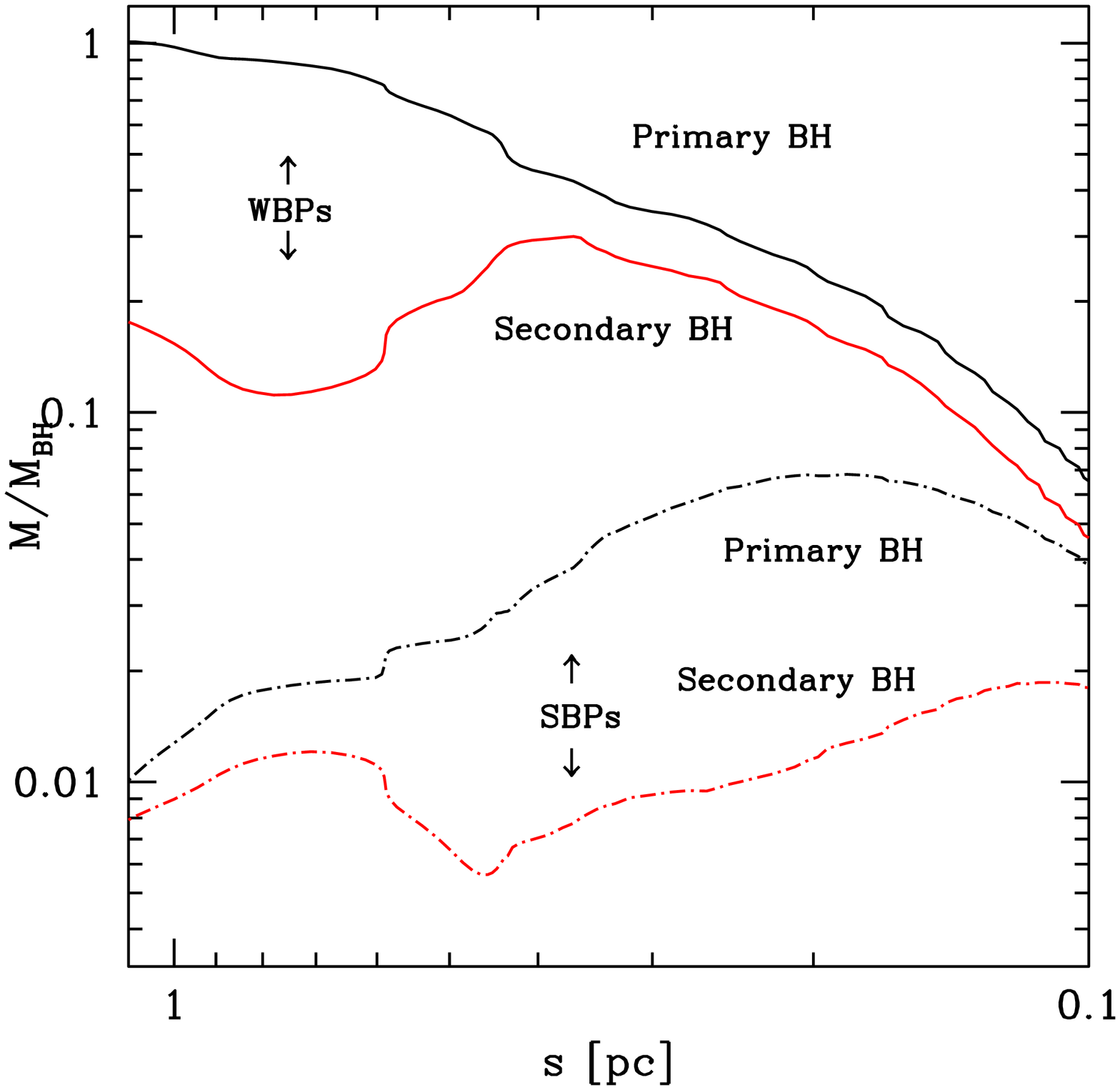} \caption{ Masses of
  bound gas particles as a function of binary separation $s$ in the
  final stages of the higher resolution simulation (between $\approx
  6$ and $\approx 8.5$ Myrs after the onset of the simulation; see
  Fig. 3).  Black lines refer to gas particles bound to the BH
  initially at rest (primary), red lines refer to the ones bound to
  the secondary BH.  Solid and dot--dashed lines refer to WB, and SB
  particles.  }
\end{figure}

The disc gas density can be as high as $10^7$ atoms cm$^{-3}$. It is
then conceivable that, at these high densities, dissipative processes
can be important, possibly reducing the gas internal (turbulent and
thermal) energy well below the values adopted in our simulations.  If
dissipative processes reduce efficiently the gaseous internal energy,
we expect that the bound gas will form a cool disc with Keplerian
angular momentum comparable to what we found in our split simulation.
We obtain for both the primary and the secondary BH a disc radius
$R_{\rm BH, disc}\lsim 0.01$pc for the SBPs, a scale more than an
order of magnitude below our best resolution limit.  Our simplified
treatment allows us to estimate a lower limit to the accretion
timescale, assuming Eddington limited accretion:
\be t_{\rm
  acc}=\frac{\epsilon}{1-\epsilon}\,\tau_{\rm Edd} \,
\ln{\left(1+\frac{M_{\rm acc}}{M_{{\rm BH,}0}}\right)}, \ee 
where $\epsilon$ is the radiative efficiency, $\tau_{\rm Edd}$ is the
Eddington time, $M_{{\rm BH,}0}$ is the initial mass of a BH, and
$M_{\rm acc}$ is the mass of the accreted gas.  Assuming
$\epsilon=0.1$, accretion can last at least $27 $ Myrs, 
and $\lsim1 $ Myrs, if the BHs accrete
all the WBPs, and SWPs, respectively.

In summary, we have found that during the orbital inspiral, the
gravitational attraction of each BH on surrounding gas particles is
such that a gaseous mass comparable to the BH mass is present inside
the BH sphere of influence and that $2\%$ of this mass binds strongly
to each single BH.  The distribution of the most bound gas particles
suggest that an active Eddington--limited accretion phase may set in,
for a time $\lsim 1$ Myr around both BHs, with luminosities $\sim
10^{44}$ erg s$^{-1}$.  The active phase could last for a longer time
($\gsim 10$ Myr), comparable to the inspiral timescale, if all the
bound mass is accreted.  This highlights the possibility of revealing
double AGN activity, on spatial scales $\lsim 10$ pc, embedded in a
starburst.  The small extension of the accretion disc around each BH
could preserve the gas against tidal perturbation form the companion
BH, and stripping until the binary reaches separations where
gravitational wave driven inspiral starts.

\begin{theacknowledgments}
M.C. acknowledges MIUR under PRIN05 for financial support.
S.K. acknowledges support by the U.S. Department of Energy through a
KIPAC Fellowship at Stanford University and the Stanford Linear
Accelerator Center.
\end{theacknowledgments}

\bibliographystyle{aipproc}   



\end{document}